\documentclass[aip,rsi,reprint,superscriptaddress]{revtex4-1}
\usepackage{graphicx}
\usepackage{amsmath}
\usepackage{color}

\begin{document}

\def\taud{\tau_\mathrm{d}}
\def\obs{_\mathrm{obs}}
\def\Tobs{T\obs}
\def\intinf{\int_{-\infty}^\infty\!\!}
\def\eqnarr#1#2{  
\renewcommand{\arraystretch}{#1}
  \setlength\arraycolsep{0ex}
  \begin{array}{rcl}
    #2
  \end{array}
}
\def\ds{\displaystyle}
\def\arreq{&{}={}&\ds }
\def\arrap{&{}\approx{}&\ds }
\def\arrdef{&{}:={}&\ds }
\def\arrdefref{&{}=:{}&\ds }
\def\arrnone{&&\ds }
\def\arrleq{&{}\leq{}&\ds }
\def\arrgeq{&{}\geq{}&\ds }
\def\arrlt{&{}<{}&\ds }
\def\arrgt{&{}>{}&\ds }
\def\arrprop{&\propto&\ds }
\def\omegam{\omega_\mathrm{m}}

\newcommand{\change}[1]{%
	{\color{blue}%
	\ensuremath{\clubsuit\!\triangleright}#1%
	\ensuremath{\triangleleft\!\clubsuit}}%
}



\title{High passive-stability diode-laser design for use in atomic-physics experiments}

\author{Eryn C. Cook}
\author{Paul J. Martin}
\author{Tobias L. Brown-Heft}
\affiliation{Department of Physics and Oregon Center for Optics, 1274 University of Oregon, Eugene, Oregon 97403-1274}
\author{Jeffrey C. Garman}
\affiliation{Technical Science Administration, 5202 University of Oregon, Eugene, Oregon 97403-5202}
\author{Daniel A. Steck}
\affiliation{Department of Physics and Oregon Center for Optics, 1274 University of Oregon, Eugene, Oregon 97403-1274}

\begin{abstract}
We present the design and performance characterization of an external-cavity diode-laser system optimized for high stability, low passive spectral linewidth, low cost, and ease of in-house assembly. 
The main cavity body is machined from a single aluminum block for robustness to temperature changes and mechanical vibrations, and features a stiff and light diffraction-grating arm to suppress low-frequency mechanical resonances.  The cavity is  vacuum-sealed, and a custom-molded silicone external housing further isolates the system from acoustic noise and temperature fluctuations.
Beam shaping, optical isolation, and fiber coupling are integrated, and the design is easily adapted to many commonly used wavelengths. Resonance data, passive-linewidth data, and passive stability characterization
of the new design demonstrate that 
its performance exceeds published specifications for commercial precision diode-laser systems.  The design is fully documented and freely available.
\end{abstract}

\maketitle

\section{Introduction\label{sec:introduction}}

Atomic physics experiments require narrow-linewidth laser sources tuned to specific atomic resonances, and tunable external-cavity diode lasers (ECDLs) are often the light source of choice as a compact, reliable, low-cost, energy-efficient option.\cite{Wieman91}  Typical spectral linewidths on the order of 1~MHz are sufficient for many laser cooling and trapping experiments, but precision spectroscopy of narrow-linewidth transitions requires narrower-linewidth
sources. Linewidths can be narrowed by servo-locking to high-finesse external reference cavities,\cite{Young99} but the requirements are less stringent and the servo is more robust when starting with an inherently narrow and stable source.

Commercial precision diode-laser systems can cost \$20,000--\$30,000; as a result, many designs for Littrow configuration ECDLs that can be built in-house already exist.\cite{MacAdam92, Arnold98, Ricci95, Vassiliev06}  In the standard Littrow configuration, the lasing cavity is formed by the rear facet of the laser diode and a diffraction grating reflecting in first order, whose angle can be changed to tune the emission wavelength.  Temperature fluctuations and acoustic vibrations that couple to the diffraction-grating mounting arm affect the cavity length and thereby change the emitted frequency, so several groups have pursued novel designs to avoid these issues.\cite{Zorabedian88, Baillard06, Papp07thesis}  A promising example is the ``cat-eye'' laser,\cite{Zorabedian88, Baillard06} which abandons the diffraction grating in favor of a tiltable interference filter as the frequency discriminating element, and the external cavity is composed of two lens-mirror ``cat-eye'' reflectors.  One model with a long (10~cm) cavity has demonstrated a passive linewidth of 155~kHz.\cite{Baillard06}  Another design machines a Littrow-configuration external cavity out of a single block of aluminum to reduce vibrational coupling to environmental noise, and the diffraction grating is mounted on an adjustable kinematic mirror mount integrated into the main cavity block.\cite{Papp07thesis}  

\begin{figure}[ht]
\centering
\includegraphics[width=9 cm]{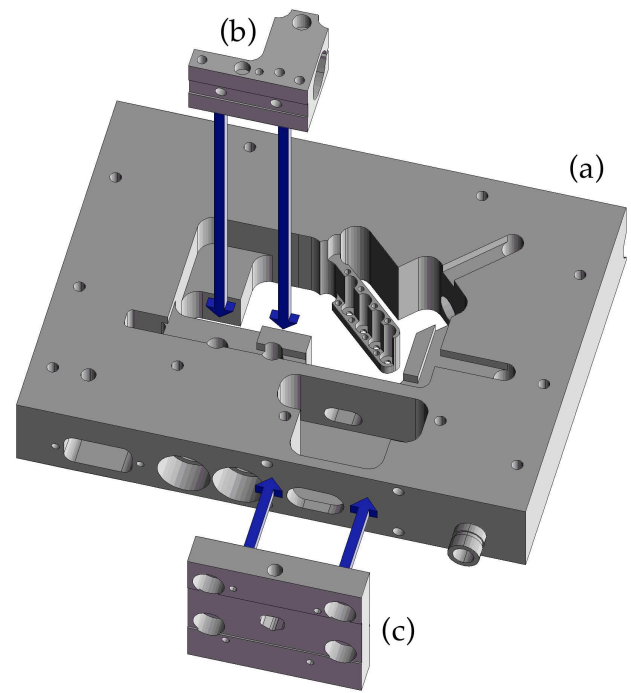}
\caption{Diagram of the (a) main laser body, (b) diode
collimation-tube mount, and (c) optical isolator mount (``aspirin tablet'' size). Not pictured are the lids and baseplate. All pieces are machined from 6061-T6 aluminum. 
\label{fig:cavity}}
\end{figure}
 
\begin{figure*}[ht]
\centering
\includegraphics[width=14 cm]{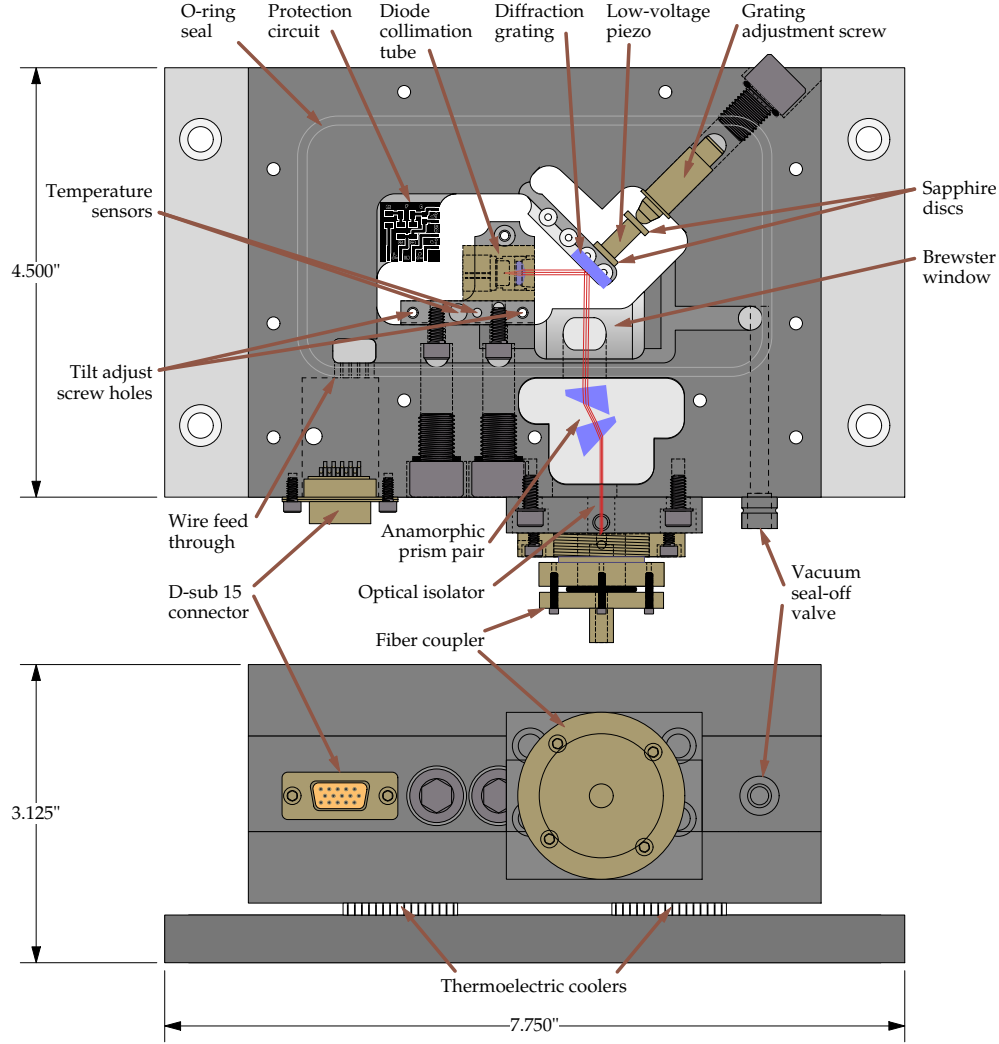}
\caption{Laser assembly: 
cutaway top view (with all peripherals but no electrical connections) and front view (with lids and baseplate).
\label{fig:diagram}}
\end{figure*}

Our design improves upon 
the basic single-block design in several ways.  In particular, the diffraction-grating arm itself is also machined as part of the main cavity block. The arm is made as stiff and light as possible to push its mechanical resonance frequency above most background noise sources.
The main cavity is hermetically/vacuum sealed to allow evacuation or backfilling with an inert gas to prevent condensation and reduce sensitivity to humidity and pressure fluctuations.  Beam shaping, optical isolation, and fiber coupling are all integrated into the assembly.  This design is readily adaptable to many different wavelengths, simply by switching the laser diode and diffraction grating (with an added shim for some wavelengths), and adjusting the beam-shaping prisms as necessary.  Gratings that would work in the geometry of our design are commercially available for many relevant atomic-physics wavelengths.
  
\section{Construction\label{sec:construction}}

The main cavity (Fig.~\ref{fig:cavity}a) is machined on a computer numerical controlled (CNC) milling
machine from a $6"\times 5"\times 1"$ block of 6061-T6 aluminum.  Use of CNC machinery
allows easy replication of the design, and complicated geometries like the diffraction-grating arm and a 
ramped surface to mount a Brewster-angle window are quickly and easily machined into the main laser body.  A pump-out port mimicking a commercial seal-off valve (DLH Industries V1021-1, to be used in conjunction with valve operator V1025-3-25) is also machined into the front face for evacuation, and all potential air traps are vented. 
Additional modules to hold the diode collimating tube (Fig.~\ref{fig:cavity}b) and the optical isolator (Fig.~\ref{fig:cavity}c) are machined separately.  The top and bottom lids are machined $0.75"$ thick, and each contains a groove for a Viton O-ring (McMaster 1201T827) to allow vacuum sealing of the 
laser cavity.  A baseplate provides mechanical support and a thermal reservoir for the thermoelectric coolers.

The main cavity body is designed to be machined from only three orientations to reduce setup labor and overall machining costs.  The diffraction-grating arm contains four 2-56 counterbored clearance holes, irregularly spaced along the length of the arm. These holes serve both to reduce the mass of the diffraction-grating arm and to provide a means to stabilize the arm during machining of main cavity body.  Additional details of the machining process are in the Appendix.

\begin{figure}[ht]
\centering
\includegraphics[width=8.5cm]{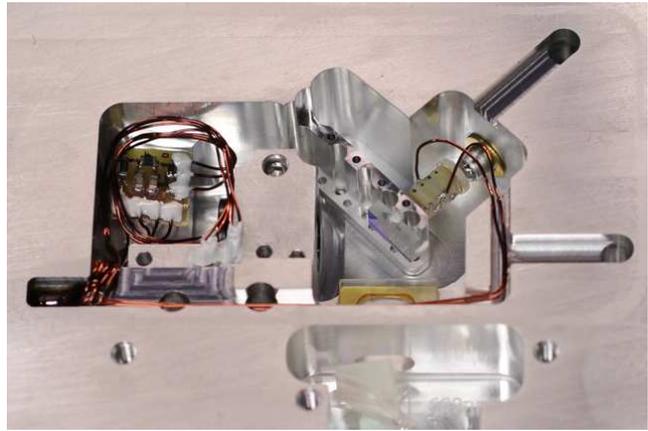}
\caption{
Assembled laser: close-in view of the main cavity of the laser with the top lid removed.
}
\label{fig:unilaser_pic}
\end{figure}

Figure~\ref{fig:diagram} shows the laser assembly with all peripherals, and Fig.~\ref{fig:unilaser_pic} shows a photograph of the laser interior, illustrating the electrical connections.  Assembly begins with a thorough (vacuum-quality) cleaning of all machined parts. Throughout construction, cleanliness is emphasized, not only to produce a sterile, vacuum-friendly environment, but also to ensure an absence of dust and foreign material. This is especially important if the laser diode has no protective housing (e.g., for antireflection-coated diodes).  Vacuum-compatible epoxy (Loctite Hysol 1C) is used to strengthen delicate solder joints and affix the diffraction grating to the arm.

The diode is held in a collimating lens tube (Thorlabs LT230P-B) which is in turn clamped in the 
collimation-tube mount (Fig.~\ref{fig:cavity}b).  This module is firmly bolted to the wall of the main cavity body, with access to the mounting bolts closed and sealed using 3/8-24 screws and Viton O-rings (McMaster 1201T23).  The face that contacts the main cavity body is lightly coated with a vacuum grease (Apiezon M) to aid in heat transfer and to help prevent long-term creep, which could otherwise occur from mechanical stresses created when the module is fixed to the main cavity body.

Prior to firmly securing the diode collimation tube module, the vertical tilt of this module can be adjusted with the help of two temporary screws. These screws have rounded ends, and are threaded vertically through the front and back of the mount, supporting the module against the bottom lid.  
By careful differential adjustment of the screws, small changes in vertical tilt can be made. As the alignment is optimized, the lasing threshold is observed until a minimum is found, and this minimum is monitored as the module is secured and the tilt-adjust screws are removed.  The vertical degree of freedom is then fixed, and the collimation-tube mount is effectively a part of the rigid cavity body: tapping and pushing on the module has no effect on cavity alignment.

The diffraction-grating mounting arm is a critical element of the laser, as its mechanical stability is a typical weak point of most diode-laser designs.  It is carved directly into the main cavity block, and designed to be as stiff and light as possible. The grating itself is custom-manufactured to a small size (6~mm high, 12~mm wide, 3~mm thick) in order to decrease the total mass of the grating arm, keeping mechanical resonance frequencies above most of the ambient acoustic noise.  Gross adjustments of the diffraction grating angle can be made using a precision 1/4-100 screw (New Focus 9376-K), with an access port sealed using another 3/8-24 screw and O-ring.  Fine wavelength adjustments are possible using a low-voltage piezo (Noliac SCMAP02/S2/A/5/5/10/60/10.6/1000) epoxied between two sapphire discs (Swiss Jewel Company W7.87) and inserted between the precision screw and the back of the grating mounting arm.  The piezo is driven by an ordinary, 15~V op-amp circuit (buffered by a high-current, unity-gain amplifier, BUF634), which typically tunes the laser $\sim\!2$~GHz/V.   A small O-ring sits in a groove in the bottom lid, compressed against the bottom of the tip of the diffraction-grating arm to provide shear damping of the arm motion.

Electrical connections enter through a D-sub 15 female connector on the front face of the laser.  
Polyimide-coated wires (MWS 23 AWG, HML insulation) pass through channels potted with a high-temperature epoxy (Epotek 353ND) before proceeding to two temperature sensors, the piezo, and the diode (through a simple protection circuit\cite{MacAdam92, Meyrath03notes}). We use a 50-k$\Omega$ thermistor for feedback control of temperature and an AD590 sensor
to drive an LCD temperature display.  Two thermoelectric coolers (Laird 56460-501) are positioned between the bottom lid and the baseplate, with a thin layer of thermal conductive paste (Arctic Silver 5) to aid heat transfer.

\begin{figure}[ht]
\centering
\includegraphics[width=8.5cm]{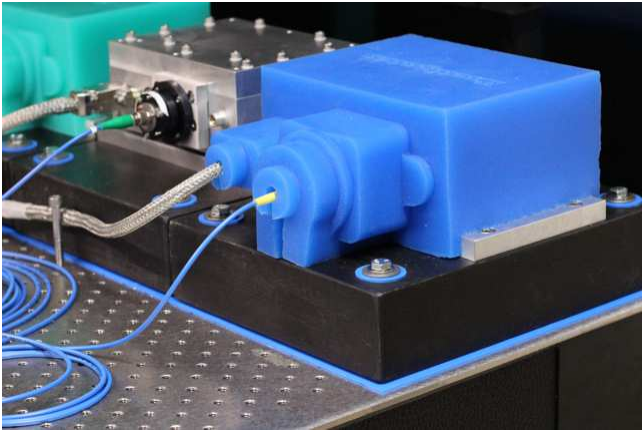}
\caption{Assembled laser mounted on table: A composite block and viscoelastic cushioning material isolate the laser from table vibrations, and a flexible, molded silicone cover offers further acoustic isolation and temperature regulation.}
\label{fig:mounting}
\end{figure}

The main cavity is sealed at the beam output by a window cut from a microscope slide. The window is glued with the thermal epoxy to a ramp cut at Brewster's angle in the main cavity body. The exiting beam then passes through an anamorphic prism pair (Thorlabs PS871-B), with the prism positions calculated to optimize the symmetry of the exiting beam and to center it on the correct output location.  Beam paths and prism locations are laser-etched onto a 0.064"-thick acrylic alignment template and fixed to the bottom of the prism cavity with five-minute epoxy (the guide can also be printed on sturdy paper).  The prisms are then attached in place on the template (after confirming proper alignment of the output beam) with small drops of five-minute epoxy.
The beam then passes through an optical isolator mounted in a separate aluminum block bolted to the main cavity body (Fig.~\ref{fig:cavity}c).  We have two versions of the isolator mount---one for small ``aspirin-tablet'' isolators (OFR  body type D) and another for larger versions (OFR body type II). Finally,  the light is output to a fiber coupler (Oz Optics HPUC-23A series) which is attached directly to the isolator mount. This minimizes the need for future adjustments of the fiber coupling and ensures that failed laser diodes can be replaced without the hassle of realigning an entire beam line.

Vibrational isolation of the laser is enhanced via additional damping components (Fig.~\ref{fig:mounting}).  For decoupling from the optical table, the assembled laser is mounted on an aggregate block (custom-molded from Castinite, with 1/4-20 threaded inserts to match the laser 
baseplate), which in turn is placed on a viscoelastic cushioning material (E-A-R Specialty Composites, Isodamp C-1002-06). A flexible silicone housing (custom-molded in-house from Smooth-On Mold Max 15T) encases the laser, reducing sensitivity to acoustic and thermal environmental perturbations.

The design described above and shown in Figs.~\ref{fig:cavity} and \ref{fig:diagram} is the 
third  generation of this general design.  To minimize resonances and push them to high frequencies, we decreased the mass of the grating arm by removing more material and using a smaller diffraction grating.  We also added a Viton O-ring under the arm for shear damping (this replaced a magnet introduced for eddy-current damping, which was ineffective), and we thickened the lids from 1/2" to 3/4" to suppress lid-dependent resonances.  We also lowered the position of the precision screw to push in the vertical center of the grating arm, after finding that some coupling between vertical tilt and the desired horizontal adjustment was leading to mode-hopping and alignment problems. The valve was not machined into the front face of the laser in the previous versions: we attempted to shrink/press-fit the commercial aluminum seal-off valve 
into a socket, but this did not seal consistently.  Finally, the original version lacked a Brewster window: the back facet of the first beam-shaping prism was glued to the exit window to create the seal, but this geometry would have made it difficult to use the same cavity body for a variety of laser wavelengths.

\section{Characterization}
Our diode-laser system is designed primarily with the goal of long-term stability in mind, but the reduction in sensitivity to acoustic perturbations ensures that the passive short-term stability and linewidth performance exceed published specifications for commercial, high-precision diode-laser sources.

The short-term stability was characterized in three ways: first, we measured the cavity resonances by performing resonance spectroscopy on the diffraction grating arm; second, we examined the power spectral density of the laser spectrum; third, we measured the delayed self-heterodyne beat note.

\begin{figure}[ht]
\centering
\includegraphics[width=8.5cm]{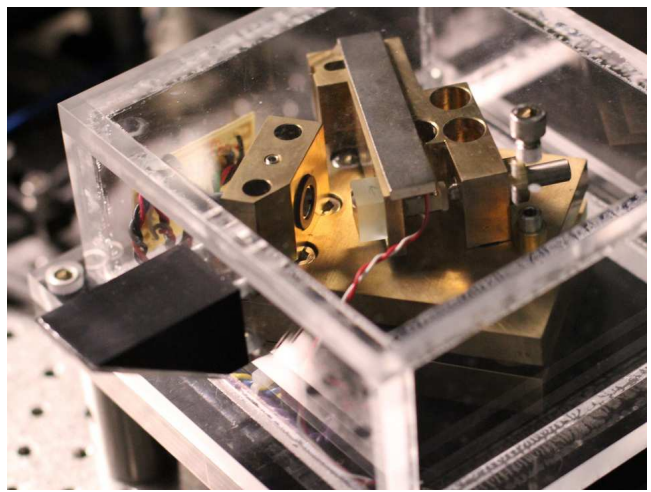}
\caption{Older ``bronze'' diode laser in use in our lab: We compare this model to the new design in all stability measurements.}
\label{fig:old780}
\end{figure}

For comparison, the same measurements were performed on an older 780-nm ECDL constructed in our lab and used to cool $^{87}$Rb (Fig.~\ref{fig:old780}). This laser is a 
descendant\cite{Steck01} of the well-known design of Ricci \textit{et al}.\cite{Ricci95}
Compared to the original design, the cavity length is increased, the diffraction angle is decreased, and the pivot of the diffraction grating arm is moved to the plane of the diode emission to optimize the mode-hop free tuning range.\cite{McNicholl85} Machined from a fatigue-resistant bronze alloy with high thermal conductivity, the laser is enclosed in an acrylic housing.  An auxiliary aluminum plate and a layer of Sorbothane are clamped directly to the aluminum baseplate for vibration damping.   Both the ``bronze'' 780-nm laser and our new test version at 780~nm use the same diode (Sharp GH0781JA2C) and collimating lens tube.

\subsection{Cavity resonance spectroscopy}
\begin{figure}[ht]
	\begin{center}
		\includegraphics[width=8.5cm]{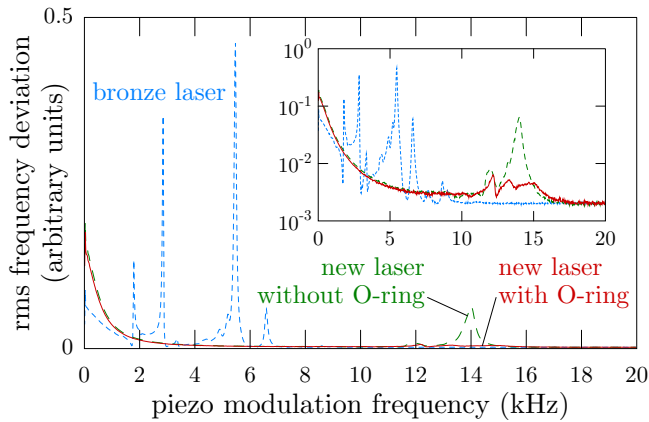}
	\end{center}
	\vspace{-3mm}
	\caption{%
Cavity resonance spectroscopy: rms error signal as a function of the driving frequency of the laser's piezo. Included are plots of the new design both with (solid red line) and without (dashed green line) shear damping as well as our lab's original bronze Littrow laser model (dotted blue line). In the inset, the same data are plotted on a logarithmic vertical scale.
	\label{fig:resonances}}
\end{figure}

We investigated the susceptibility of the cavity to acoustic noise by the measuring the resonances of the most vibration-susceptible and critical element
for frequency stability---the diffraction grating arm.  An external Fabry--Perot cavity was slowly servo-locked (100-ms integration time constant) to keep the laser near the half-maximum point of the cavity resonance.  A small-amplitude (e.g., 0.03~V$_\mathrm{p-p}$ through a 20.2-dB attenuator) chirped sine wave was swept from 0 to 20~kHz over 20~s and applied to the laser tuning piezo. We monitored the error signal from the Fabry--Perot servo and plot the root-mean-square (rms) frequency excursions of the laser averaged over 200 sweeps in Fig.~\ref{fig:resonances}. 

For comparison, we have included the same resonance measurements performed on the 
bronze design described above.  At low frequencies, the modulation simply displaces the grating, tuning the laser directly; since this tuning response varies with design (the diffraction-grating arm mass, stiffness, material, and incident angle are all different), we scale the rms 
excursions so that the high-frequency noise floors (which fall off at 
6~dB/octave, as expected for a $1/f^2$ driven response) are consistent.

Due to the stiff, lightweight grating arm, 
we do not observe any significant resonances below around 12 kHz in our new design, and the addition of the damping
O-ring beneath the end of the grating arm effectively reduces the amplitude of the fundamental resonance without hampering the ability to tune the laser at low frequencies.  
We confirmed that the visible peaks are associated with the grating arm by attaching various weights to the arm and observing the shifts of the resonances.  In some cases we have also observed peaks at 3.5 and 7~kHz whose amplitudes are nonlinear as a function of drive amplitude. These are evidence of an improperly secured grating and can be eliminated by using a sufficiently strong epoxy bond. 

The previous version of our new design had a larger diffraction grating and 
more mass in the grating arm, as described above; we found that decreasing the grating arm mass and inserting the O-ring under the arm for shear damping helped to minimize resonances and push them higher in frequency.  By reducing the mass of the arm and grating, we were able to increase the frequency of the lowest resonance by nearly 30\%.  We also observed an apparent coupling between the arm resonances and the lids in the previous version, which was eliminated by increasing the lid thickness. 

\subsection{Frequency-noise spectrum}

Using a digital spectrum analyzer, we monitored the transmission of a 780-nm laser through a Rb vapor cell to measure the frequency-noise spectrum.  A slow, integrating servo (3~s time constant) maintained the laser frequency near the half-maximum point of a saturated-absorption feature ($F~{=}~3\longrightarrow F'~{=}~3,4$ hyperfine crossover) of
$^{85}$Rb.  The intensity fluctuations of the detected light are converted to frequency noise using the
well-known hyperfine structure.

We observed that the current controller used to power the laser had the largest effect on the overall baseline of the frequency-noise spectrum. For most lasers in our lab we use a home-built supply\cite{Meyrath03notes} with $\sim\!\!10$-$\mu$A rms noise,
but initial measurements demonstrated that one feature observed near 700~kHz could be identified as arising from noise in the current supply.  Low-pass filtering the current supply output decreased the frequency noise significantly, and a modification of the circuit design to increase the 
sense-resistor value and reduce gain in the feedback loop further improved the characteristics.  A commercial (Vescent Photonics D2-105) version of the Libbrecht-Hall current controller\cite{Libbrecht93} offers superior noise performance (100-nA rms noise), and was used to power the lasers during all measurements presented here (except for the heterodyne measurement in Fig.~\ref{fig:780het}). 

\begin{figure}[ht]
	\begin{center}
		\centering
		\includegraphics[width=8.5cm]{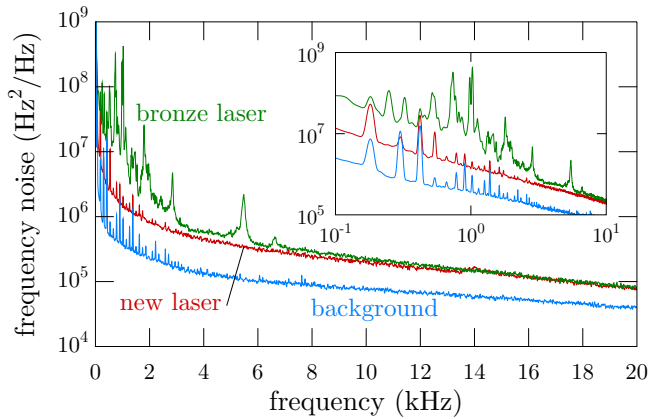}
	\end{center}
	\vspace{-3mm}
	\caption{%
		Frequency-noise spectral density for the old bronze design (green), new stable design (red), and detector background (blue). Inset are the same data on a logarithmic scale from 100~Hz to 10~kHz.
	\label{fig:PSD}}
\end{figure}

Figure~\ref{fig:PSD} shows the frequency-noise spectra for the two lasers, as well as a background noise floor (measured with the detector blocked), with each curve averaged for several minutes. Many features can be identified as electrical artifacts that appear in the background as well, but for the older bronze design, there are some peaks that correspond to the grating arm resonances in Fig.~\ref{fig:resonances} (e.g., at 1.8, 2.9, 5.5, and 6.6~kHz), while the $\sim\!\!14$~kHz grating arm resonance of the new design is barely visible.  No features were observed in the 20--100~kHz band in the spectrum of either laser.  The high-frequency baselines were similar for both the bronze laser and the new design, but the new design was much quieter in the acoustic (0--20~kHz) band.

Integration of the frequency-noise spectral density gives a measure of the laser's
rms frequency noise $\Delta\nu_{\rm{rms}}$ within a particular frequency band.  We compute
\begin{equation}
\Delta\nu_{\rm{rms}}~{=}~\eta\sqrt{R\int_{f_1}^{f_2}\overline{P}(f)\,df}
\end{equation}
where $\overline{P}(f)$ is the spectrum analyzer output in 
watts (normalized by subtracting the background), $R$ is the 50~$\Omega$ internal impedance of the spectrum analyzer, and $\eta$ is the voltage-to-frequency conversion factor derived from the hyperfine spectrum.
Over the 20~Hz--100~kHz band, we find $\Delta\nu_{\rm{rms}}~{=}~296(26)$~kHz for the bronze laser and 146(8)~kHz for the new design (the lower 20-Hz cutoff was chosen to remove the contribution of the dc bias in the measurement, which produces a sharp dc peak broadened by the resolution bandwidth  of the analyzer).  The measured contribution from noise in the 20--100~kHz band was negligible for both lasers.   The contribution from noise above 100~kHz, computed by extrapolating the high-frequency baseline of the spectrum, also has a negligible effect (though this ignores the contribution of relaxation oscillations in the GHz range).

As an extreme test of the lasers' response to environmental perturbations, we also subjected 
them to a harsh environment by sounding an air horn from across the optical table (a distance of 
4~m) whose intensity in each laser's vicinity was 105 dBA. The new laser responded very little, and we were able to use the same $F~{=}~3\longrightarrow F'~{=}~3,4$ saturated-absorption feature as a frequency reference.  However, the bronze laser was quite sensitive to the air horn, exhibiting frequency excursions that far exceeded the width of this feature.  Instead we used the side of the Doppler-broadened $F=3\longrightarrow F'$ collective resonance of $^{85}$Rb; this gave a reduced sensitivity, allowing us to measure the bronze laser's response. The reduced sensitivity caused the unperturbed laser to be indistinguishable from the noise floor, and leads to a high-frequency floor which (after conversion to a frequency-noise density) is offset from the floor of the crossover reference. We 
attempted to use the same, Doppler-broadened feature with the new laser, but did not observe a significant response to the perturbation.

\begin{figure}[ht]
	\begin{center}
		\includegraphics[width=8.5cm]{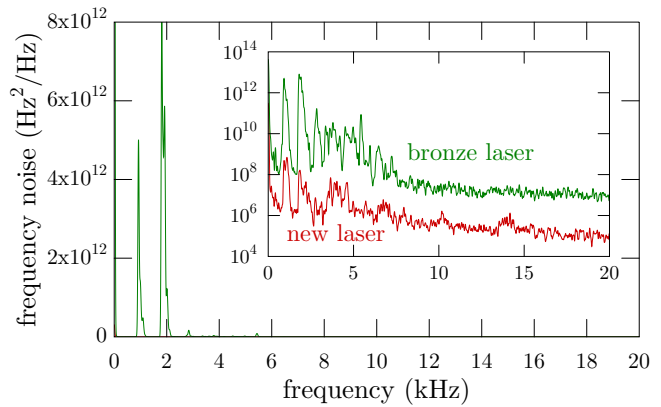}
	\end{center}
	\vspace{-3mm}
	\caption{%
Spectral response exhibited by the old bronze design (green) and the new stable design (red) 
for an air-horn perturbation (105~dBA local sound intensity).  In the inset, the same data are plotted on a logarithmic vertical scale.
	\label{fig:airhorn}}
\end{figure}

Figure~\ref{fig:airhorn} shows the laser frequency-noise spectrum as the air horn is sounded. Sensitivity to the perturbation is reduced in our new design by two orders of magnitude. The response of the $\sim\!\!14$~kHz grating-arm resonance can be seen, along with lower-frequency peaks, many of which are common to both lasers, and therefore reflect either details of the air-horn spectrum or acoustic room resonances.  To quantify the effect of the perturbation, we measure $\Delta\nu_{\rm{rms}}~\sim~1$~MHz for the new design, versus 40~MHz for the old design (an increase by factors of 7 and 135 relative to the non-perturbed values, respectively).

During these measurements, we also used a more robust servo (with both proportional and integral feedback) to lock each laser explicitly to the crossover transition. We observed that common laboratory perturbations (e.g., bumping the optical table or laser and dropping ball drivers or Allen keys on the table) were occasionally catastrophic to the bronze laser's stability, but had little effect on our new design.

\subsection{Delayed self-heterodyne linewidth characterization}
 We characterized the lasers' short-term spectral width via the standard self-heterodyne 
technique,\cite{Gallion82, Gallion84, Mercer91, Ludvigsen98} where 
the laser output 
splits into two components---one with a frequency shift, and one with a time delay---which
beat on a detector. The
resulting signal is recorded through a spectrum analyzer.  
Our setup used delays in the range of 
$\sim\!\!11$--44~$\mu$s, corresponding to 2--9~km of SMF-28 fiber, 
and a nominal frequency shift of $\omegam/2\pi~{=}~80$~MHz, using first-order diffraction by an acousto-optic modulator.

To model the spectrum, we assume 
a frequency-noise spectrum consisting of white-noise and $1/f$-noise components,\cite{Mercer91} 
parameterized by $\gamma$ and $k$, respectively:
\begin{equation}
  S_\omega(\omega)~{=}~\gamma + \frac{k}{\omega}.
\end{equation}
Here, $S_\omega(\omega)$ is the one-sided spectral density of frequency fluctuations, which we may
define in terms of the laser phase $\phi(t)$ by
\begin{equation}
  S_\omega(\omega):=\intinf \left\langle{\dot{\phi}(t)\,\dot{\phi}(t+\tau)} \right\rangle\cos\omega\tau\,d\tau .
\end{equation}

With this noise model, the normalized spectrum from the spectrum analyzer is given 
by\cite{Mercer91}
\begin{align}
    s_\mathrm{\scriptscriptstyle{SH}}(\omega) 
    =  \frac{1}{2\pi}\intinf d\tau \, & \Big\{ \,\,\,e^{i\Delta\tau-\gamma\min\{|\tau|,|\taud|\}}     
    \label{bigfitfunc}\\
    &\times |\tau|^{k\tau^2/\pi}
    |\tau+\taud|^{-k(\tau+\taud)^2/2\pi}\nonumber\\
    &\times|\taud|^{k\taud^2/\pi}
    |\tau-\taud|^{-k(\tau-\taud)^2/2\pi}\nonumber\\
    &\times e^{-(\delta\omega_\mathrm{\scriptscriptstyle RB}\tau)^2/16\log 2
    -\delta\omega_\mathrm{drift} \tau/2} \,\,\Big\}, \nonumber
\end{align}
where $\Delta:=\omega-\omegam$, and $\taud$ is the optical-fiber time delay.
We have 
included two cutoffs in the integrand corresponding to the Gaussian resolution bandwidth
$\delta\omega_\mathrm{\scriptscriptstyle RB}$ (full width at half maximum) of the spectrum analyzer,
and also a phenomenological exponential cutoff $\delta\omega_\mathrm{drift}$ to model
drifts of $\omegam$ throughout the acquisition and averaging of the spectra over many minutes.
These cutoffs convolve the expected self-heterodyne spectrum with Gaussian and Lorentzian
functions, respectively. The resolution bandwidth 
$\delta\omega_\mathrm{\scriptscriptstyle RB}/2\pi$ is nominally 1~kHz, and
$\delta\omega_\mathrm{drift}$ is of the same order or much smaller.

We measure the self-heterodyne spectra with multiple delays, and fit them to the the above spectral model,
taking $\gamma$, $k$, $\delta\omega_\mathrm{\scriptscriptstyle RB}$,
and
$\delta\omega_\mathrm{drift}$ to be fitting parameters.  We also take the overall scale and an overall
offset (corresponding to a detector noise floor of typically $-92$~dBm) as fitting parameters,
and we 
allow for an extra fitting parameter, the amplitude of a 
coherent ``spike'' (delta function) at $\Delta=0$ (in addition to any expected in the spectrum),
which is of course also broadened by 
 $\delta\omega_\mathrm{\scriptscriptstyle RB}$ and
$\delta\omega_\mathrm{drift}$.
Thus,  $\delta\omega_\mathrm{\scriptscriptstyle RB}$,
$\delta\omega_\mathrm{drift}$, and the last amplitude serve to fit the center spike, while the other
parameters characterize the spectral tails.
The fiber delay $\taud$ is also a fitting parameter, but only for the shortest fiber delays
where delay-dependent fringes are clearly visible in the spectral tails. This allows us to properly account for the wavelength-dependent fiber delay.

Once the $\gamma$ and $k$ parameters of the frequency-noise spectrum are determined, inferring the width of the
laser spectral line is somewhat subtle.  While the $\gamma$-dependent white noise leads to 
a Lorentzian spectrum, the $k$-dependent, $1/f$ noise leads to a divergent spectral width.
Thus, the spectral width must be computed with respect to an observation time\cite{Cutler66}
$\Tobs$, 
since the spectral width diverges with $\Tobs$.  (The spectral width also diverges
as $\Tobs\longrightarrow 0$ owing to time-frequency uncertainty.)
To compute the spectral width, we thus numerically evaluate an 
integral expression\cite{Steck06notes} analogous to Eq.~(\ref{bigfitfunc}) for the laser spectrum, incorporating the observation time, and 
compute its full width at half maximum.
This can be computed for uniform observation of the signal over a time $\Tobs$, or a Gaussian-weighted
observation in time, corresponding to a Gaussian resolution bandwidth of a
(hypothetical) spectrum analyzer for the optical frequencies.

While the frequency-noise spectrum characterization could only be performed for 780-nm model diode lasers due to the availability of the Rb vapor cells for calibration, we measured the delayed self-heterodyne signal for both 780-nm and 922-nm diode lasers, as well as for two diode lasers at 689~nm, one with the usual 
2~cm cavity length and one with a longer, 10~cm cavity (otherwise identical to the 2~cm models).  Table~\ref{lasers_built} summarizes the laser parameters and measurement results.

\begin{figure*}[tb]
	\begin{center}
		\includegraphics[scale=1]{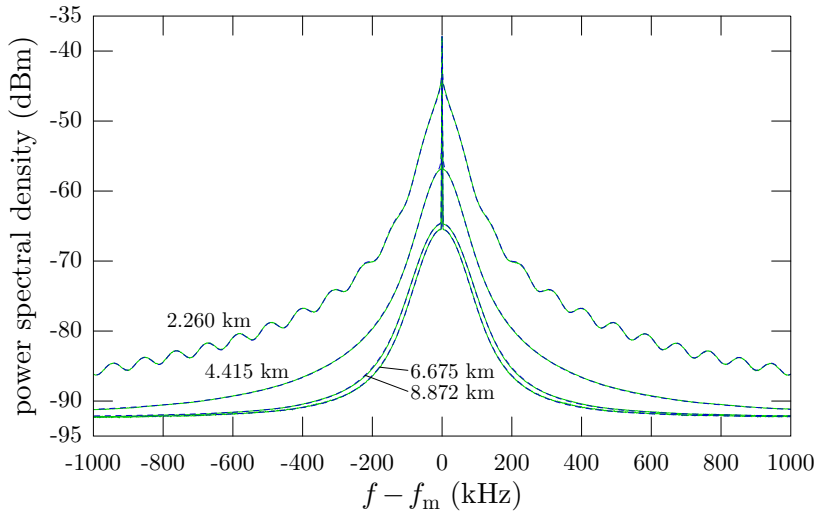}
	\end{center}
	\vspace{-3mm}
	\caption{%
		Measured self-heterodyne spectra (solid green lines) and fits (dashed blue lines)
		to the model described in the text, for four different fiber-optic delay lengths, for the 
		922~nm laser. 
		Averaged over the four fits, the fitted spectral parameters are 
		$\gamma/2\pi~{=}~1.86(29)$~kHz and $\smash{\sqrt{k}/2\pi}~{=}~27.0(15)$ kHz.
	\label{fig:922fit}}
\end{figure*}

The measured spectra for the 922-nm laser are plotted in
Fig.~\ref{fig:922fit} with the fitted spectra for four different delays.
The shortest (2.260~km) delay line gives a fitted delay of 11.116~$\mu$s;
the longest delay (8.872~km) thus corresponds to a 43.637-$\mu$s delay.
Averaging the fitted parameters from the four spectra, we find a white-noise width of
$\gamma/2\pi~{=}~1.86(29)$~kHz, and a $1/f$-noise parameter of $\smash{\sqrt{k}/2\pi}~{=}~ 27.0(15)$~kHz.
These parameters give spectral widths of
19, 32, and 75 kHz for $\Tobs~{=}~50$, 100, and
1,000~$\mu$s, respectively.
Alternately, the spectral width is 40 or 71 kHz for a Gaussian resolution bandwidth of 
$\delta\omega/2\pi~{=}~10$
or 1~kHz, respectively.

The measured spectra for the 780-nm models indicate a white-noise width of $\gamma/2\pi~{=}~1.8(3)$~kHz for the new laser and 15(1)~kHz for the older bronze laser, and $1/f$-noise parameters of 
$\smash{\sqrt{k}}/2\pi~{=}~$
93.3(5) and 85(2)~kHz, respectively.  While the white-noise width of the new 780-nm laser was substantially narrower than the older bronze laser, and in fact was similar to that found for the new 922-nm laser, the $1/f$ parameter dominates and caused the two lasers to have 
similar overall spectral widths of $\sim\!200$~kHz for $\Tobs~{=}~100~\mu$s.  The narrower white-noise width of the new design is due partially to improved mechanical stability, but increased diffraction-grating resolution probably dominates.
The new design uses a grating with 1800 lines/mm, compared to 1200 lines/mm in the old bronze laser, and the incident angle is greater (45$^\circ$ versus 28$^\circ$), resulting in a larger illuminated grating area.

\begin{figure}[ht]
	\begin{center}
		\includegraphics[width=8.5cm]{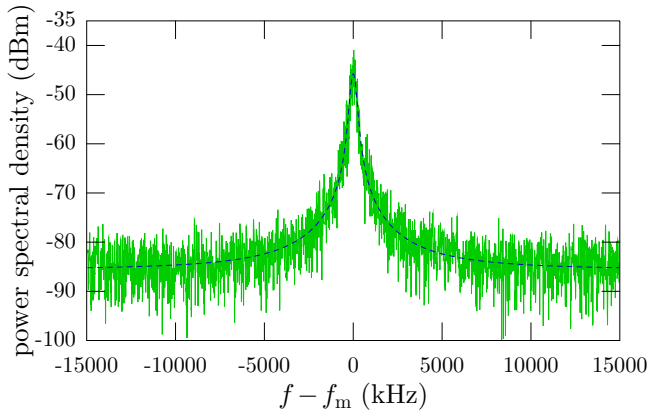}
	\end{center}
	\vspace{-3mm}
	\caption{Heterodyne signal (solid green line) and fit (dashed blue line) between two identical 780~nm lasers, acquired using a 20-kHz Gaussian resolution bandwidth. The fitted spectral parameters are $\gamma/2\pi~{=}~20(10)$~kHz and $\smash{\sqrt{k}/2\pi}~{=}~90(40)$ kHz.
	\label{fig:780het}}
\end{figure}

For comparison, we also examined the heterodyne spectrum (acquired with a 20-kHz Gaussian resolution bandwidth) between two identical new-model 780-nm lasers (Fig.~\ref{fig:780het}), with one laser powered by the Vescent current controller and the other by the improved, filtered home-built supply.  We fit the data with the heterodyne analog of Eq.~(\ref{bigfitfunc}),
and infer linewidth parameters of $\gamma/2\pi~{=}~20(10)$~kHz and $\smash{\sqrt{k}}/2\pi~{=}~$90(40)~kHz.  The uncertainty in these values is large due to the fact that averaging the heterodyne spectrum of free-running lasers is problematic, but the overall spectral width of $\sim\!\!200$~kHz for $\Tobs~{=}~100~\mu$s is consistent within error with the self-heterodyne measurement.

The measured spectra for the 689-nm model with the same short 2~cm cavity length 
indicates a linewidth  wider than either the 780 or 922-nm models: we find a white-noise width of $\gamma/2\pi~{=}~$93(36)~kHz and a $1/f$-noise parameter of $\smash{\sqrt{k}/2\pi}~{=}~$87.0(7) kHz, leading to an overall spectral width of 254~kHz for $\Tobs~{=}~100~\mu$s.  Lengthening the cavity did, however, decrease the linewidth significantly: for the 10~cm cavity length, we find a white-noise width of $\gamma/2\pi~{=}~$4(1)~kHz and a $1/f$-noise parameter of $\smash{\sqrt{k}/2\pi}~{=}~$22(1) kHz, leading to an overall spectral width of 11.7~kHz for $\Tobs~{=}~100~\mu$s. 

The observed linewidth reduction due to the cavity-length extension is consistent with theory. The Schawlow-Townes linewidth is proportional to the cavity bandwidth squared,\cite{Schawlow58} so we expect the white-noise linewidths of the two lasers (which use identical diodes and have negligible front-facet reflectivity) to be related:
\begin{equation}
\frac{\gamma_{\rm{short}}}{\gamma_{\rm{long}}} = \left(\frac{nd+L_{\rm{long}}}{nd+L_{\rm{short}}}\right)^2,
\end{equation}
where $n$ is the semiconductor index of refraction, $d$ is the length of the solitary diode, and the $L_i$ are the lengths of the passive, extended cavities.
For typical diode parameters ({$nd\sim600~\mu$m}) and our external cavity lengths, we find that the measured linewidths are consistent with this relation.  Furthermore, the extended length reduces the sensitivity of the laser frequency to changes in diode index of refraction, and the above linewidth reduction is also equal to the square of the reduction in the frequency chirp $\partial\nu/\partial I$.\cite{Kazarinov87} We saw evidence of this effect, measuring current tuning values of 57(15) and 15(3)~MHz/mA for the short and long lasers, respectively, where the uncertainty
corresponds to systematic variations that depend on the overall current value and proximity to mode hops.

\begin{table*}
\begin{tabular}{|c|c||c|c||c|c||c|c|c|}
\hline
\multicolumn{2}{|c||}{Diode} & \multicolumn{2}{|c||}{Grating} & \multicolumn{2}{|c||}{Cavity} &  \multicolumn{3} {|c|}{Linewidth Parameters (kHz)}\\
\hline
\hline
\ \ wavelength \ \ & $R_{\rm{d}}$ & \ \ \ lines/mm\ \ \  & $R_{\rm{g}}$ & \ \ $L_{\rm{c}}$ (cm) \ \ & \ \ Design \ \ & \ \ \ \ \ $\gamma/2\pi$ \ \ \ \ \ & $\ \ \ \ \sqrt{k}/2\pi$\ \ \ \  &\ \ \ FWHM\ \ \  \\
 \hline
\ \ \ 780 nm \ \ \    &\ \ \  N/A \ \ \ & 1200 & \ \ 0.45 \ \ & 2.5 & old & 15(1) & 85(2) & 200\\
\ \ \ 780 nm \ \ \    &\ \ \  N/A \ \ \ & 1800 & \ \ 0.25 \ \ & 2.2 & new & 1.8(3) & 93.3(5) & 200\\

922 nm  &  10$^{-6}$  &  1500    &  0.19 & 2.2 & new & 1.86(29) & 27.0(15) & 32 \\
689 nm & 10$^{-5}$    & 2000   & 0.16  & 2.2 & new & 93(36) & 87.0(7) & 254 \\
689 nm & 10$^{-5}$    & 2000   & 0.16  & 10.0 & new & 4(1) & 22(1) & 11.7 \\
\hline
\end{tabular}
\caption{Summary of the lasers characterized by the self-heterodyne technique. $R_{\rm{d}}$ is the diode front-facet reflectivity, $R_{\rm{g}}$ is the measured diffraction grating efficiency, and $L_{\rm{c}}$ is the external cavity length. Also included are measured linewidth parameters (white-noise and $1/f$ components) and calculated FWHM for $\Tobs$~{=}~100~$\mu$s. The diode and diffraction grating combination used for each wavelength are, respectively: old 780~nm---Sharp Microelectronics GH0781JA2C and Edmund NT43-751; new 780~nm---Sharp Microelectronics GH0781JA2C and Richardson 53-*-330H; 922~nm---Sacher SAL-920-60 and Richardson 53-*-239H; and 689~nm---Sacher SAL-690-25 and Richardson 53-*-059H.  
\label{lasers_built}}
\end{table*}

There are several reasons why different diodes used in the same short laser cavity might display very different linewidths.  As a result of coupling between phase and intensity noise, the fundamental quantum Schawlow-Townes linewidth is known to be broadened in semiconductor lasers by a factor\cite{Henry82} that typically ranges from \mbox{3--7}.  While this factor is not often quoted by laser-diode manufacturers, since the technical $1/f$ noise component typically dominates ultimate linewidths, it could account for the increased white-noise parameter $\gamma$ observed in the 689-nm diode lasers.

Another factor to consider is that the diode-front-facet and diffraction-grating reflectivities for each model were very different:  The 780-nm diode lasers used diodes without a special front-facet anti-reflection (AR) coating, while the 922-nm and 689-nm diodes were AR-coated (with reflectivities $\le\!\!10^{-6}$ and $\le\!\!10^{-5}$, respectively).  The grating reflectivities were measured to be 0.45, 0.25, 0.19, and 0.16, respectively, for the gratings used in the old 780-nm model, the new 780-nm model, and the 922-nm and 689-nm models.  Although previous groups have found that grating resolution is a much more important factor than grating or front-facet reflectivities,\cite{Loh06, Saliba09} the relative resolution (ratio of wavelength to groove separation) of all of our gratings are similar in the new design, although the resolution of the old 780-nm model is lower.  However, the different reflectivity parameters suggest that competition between the external cavity modes and those of the diode itself may play a role in increasing the $1/f$-noise parameters of the new 780-nm and 689-nm models.

\section{Conclusion}\label{sec:conclusion}

We have constructed and characterised five
lasers using this new single-block design: two prototypes at 780~nm for comparison to existing homebuilt diode lasers, two more with the same cavity length at 922 and 689 nm, and one extra-long cavity (10-cm cavity length, as opposed to 2~cm for the shorter design) for a narrower passive linewidth (see Table~\ref{lasers_built} for a list of laser parameters and linewidth-measurement results).  In addition, we have also constructed two other lasers at 994 and 914 nm to complete the set of lasers needed for precision measurements using magneto-optically trapped, neutral strontium.   
The need for so many different wavelengths, and for one ultra-narrow laser to probe strontium's ``forbidden''  $5\mathrm{s}^2\,{^1}\mathrm{S}_0\longrightarrow 5\mathrm{p}\,{^3}\mathrm{P}_1$
``clock'' transition at 689~nm, motivated our interest in designing a high-passive-stability diode-laser design adaptable to a wide range of diodes and cavity lengths.  

Our diode laser design offers passive stability and linewidths superior to published specifications of commercial precision diode-laser systems.  
For an observation time of $\Tobs~{=}~100~\mu$s, we measured linewidths of 32~kHz for an AR-coated diode at 922~nm, and $\sim$200~kHz for a non-AR coated diode at 780 nm.  
For AR-coated diodes at 689~nm in the regular and the extra-long cavities, we measured linewidths of 254~kHz and 11.7~kHz, respectively.  We have also demonstrated that our laser exhibits no low-frequency mechanical resonances and is extremely insensitive to acoustic noise.  

For comparison, the New Focus flagship ECDL, the Vortex II (TLB-6900 series), displays several resonances under 10~kHz (with the largest feature located at $\sim $500 Hz), and specifies
a linewidth of $\le\!\!300$~kHz from a simple full-width-half-max (FWHM) measurement of the heterodyne spectrum.\cite{NewFocus11brochure}  The Toptica DL Pro does not provide mechanical-vibration measurements, and specifies a 100~kHz typical linewidth, associated with a 5~$\mu$s time figure whose meaning is not stated.\cite{Toptica11brochure} Toptica also presents a Lorentzian fit to a heterodyne spectrum for two of these lasers, with a FWHM of 300~kHz, deconvolved to $150$~kHz;\cite{Toptica11cat} however, the spectrum itself is not Lorentzian, and this figure underestimates the deconvolved linewidth by a significant margin.  Vescent Photonics also uses a Lorentzian fit to characterize a non-Lorentzian spectrum, similarly underestimating to specify a fast linewidth on the order of 50 kHz for their Chip External-Cavity Laser (CECL).\cite{VescentCECL, VescentCECL2} Finally, Sacher's Lynx ECDL has a typical linewidth of 1~MHz for $\Tobs~{=}~1~$ms.\cite{SacherLynx}

A set of components for a diode-laser system using our design costs under 
\$6,000. Machining, materials, and a set of peripherals (including diffraction grating and fiber coupler) 
 costs under \$3,000.  In addition, an appropriate diode and optical isolator must be purchased; costs for these components range from \$1,000--\$3,000, depending on diode wavelength, AR-coating, isolator model, etc.  In our lab this past summer, a talented undergraduate student---who had no prior laser-assembly experience---assembled and aligned four laser systems in under a month.
All machine drawings and CAD files, detailed assembly instructions, and all circuit schematics and board layouts for the relevant electronics are available on our project web site.\cite{Unilaser_web}

\appendix*
\section{Machining notes}
While the fabrication of most of the components of the laser is
straightforward, the main cavity body is complex and requires special attention.  Here we provide an overview of the most important machining details for this component.

Machining for this project was performed with a 1993 Bridgeport Discovery 308 3-axis Vertical Machining Center (VMC).  This version of the Discovery 308 has a 5 HP max, 6,000 RPM spindle and a BT 30 taper spindle. Some deep-pocketing operations are necessary in this design (e.g., for the electrical feedthroughs) and extra care was taken in choosing path parameters, cutting speeds, and feed rates
to accommodate the reduced rigidity caused by the high tool-length-to-diameter ratio.  Corners for pockets and places where web thicknesses are critical were 
drilled prior to the roughing and finishing paths.  To minimize the number of tool changes required, and to allow for adjustability using tool-table offsets, all of the counterbore and O-ring features for the cavity-sealing 3/8-24 screws were interpolated with an end mill. 

The diffraction-grating mounting arm is designed to include four 
counterbored clearance holes for 2-56 machine screws along its length. These perform the dual purpose of 
reducing mass and providing a method for supporting the arm during final finish-pass machining of all inner walls in the main cavity pocket. These through-holes, and the holes that define the grating-arm flexure, are drilled before cavity material is removed. Initially, the main cavity is roughly machined with a 3/8"-diameter end mill, and a $\sim\!\!1$/4" web of material is left to connect the end of the diffraction grating arm to the rest of the body. A fixture plate similar to the lids is affixed to the bottom side of the main cavity via the 8-32 holes around its perimeter. The plate contains four 2-56 tapped holes, used to stabilize the arm during final-pass machining.  With the fixture plate in place, the inner-pocket walls and diffraction-grating arm are finished with a 1" long, 4 flute, 1/4" diameter end mill.  

In the same setup the ramp for the Brewster output window is machined with a bull nose cutter and a one-way parallel 3D surfacing tool path.  For a 1/4" end mill, a radial stepover of 0.010" quickly produces a reasonable finish with minimal cusp height, and still allows for a leak-tight seal when the output window is epoxied in place.  The front-face pump-out port has an O-ring feature that is interpolated using a 3/4"-diameter, 1/32"-convex-radius end mill (MSC 03017498). 

Construction of a complete set of the components detailed in Fig.~\ref{fig:cavity} above requires about 20 hours of (professional) machining time.  The University of Oregon Technical Science Administration machine shop can fabricate these for other institutions; contact the authors for more information.


\begin{thebibliography}{31}
\expandafter\ifx\csname natexlab\endcsname\relax\def\natexlab#1{#1}\fi
\expandafter\ifx\csname bibnamefont\endcsname\relax
  \def\bibnamefont#1{#1}\fi
\expandafter\ifx\csname bibfnamefont\endcsname\relax
  \def\bibfnamefont#1{#1}\fi
\expandafter\ifx\csname citenamefont\endcsname\relax
  \def\citenamefont#1{#1}\fi
\expandafter\ifx\csname url\endcsname\relax
  \def\url#1{\texttt{#1}}\fi
\expandafter\ifx\csname urlprefix\endcsname\relax\def\urlprefix{URL }\fi
\providecommand{\bibinfo}[2]{#2}
\providecommand{\eprint}[2][]{\url{#2}}

\bibitem[{\citenamefont{Wieman and Hollberg}(1991)}]{Wieman91}
\bibinfo{author}{\bibfnamefont{C.~E.} \bibnamefont{Wieman}} \bibnamefont{and}
  \bibinfo{author}{\bibfnamefont{L.}~\bibnamefont{Hollberg}},
  \bibinfo{journal}{{Rev.\ Sci.\ Instrum.}} \textbf{\bibinfo{volume}{62}},
  \bibinfo{pages}{1} (\bibinfo{year}{1991}).

\bibitem[{\citenamefont{Young et~al.}(1999)\citenamefont{Young, Cruz, Itano,
  and Bergquist}}]{Young99}
\bibinfo{author}{\bibfnamefont{B.~C.} \bibnamefont{Young}},
  \bibinfo{author}{\bibfnamefont{F.~C.} \bibnamefont{Cruz}},
  \bibinfo{author}{\bibfnamefont{W.~M.} \bibnamefont{Itano}}, \bibnamefont{and}
  \bibinfo{author}{\bibfnamefont{J.~C.} \bibnamefont{Bergquist}},
  \bibinfo{journal}{{Phys.\ Rev.\ Lett.}} \textbf{\bibinfo{volume}{82}},
  \bibinfo{pages}{3799} (\bibinfo{year}{1999}).

\bibitem[{\citenamefont{MacAdam et~al.}(1992)\citenamefont{MacAdam, Steinbach,
  and Wieman}}]{MacAdam92}
\bibinfo{author}{\bibfnamefont{K.~B.} \bibnamefont{MacAdam}},
  \bibinfo{author}{\bibfnamefont{A.}~\bibnamefont{Steinbach}},
  \bibnamefont{and} \bibinfo{author}{\bibfnamefont{C.}~\bibnamefont{Wieman}},
  \bibinfo{journal}{{Am.\ J.\ Phys.}} \textbf{\bibinfo{volume}{60}},
  \bibinfo{pages}{1098} (\bibinfo{year}{1992}).

\bibitem[{\citenamefont{Arnold et~al.}(1998)\citenamefont{Arnold, Wilson, and
  Boshier}}]{Arnold98}
\bibinfo{author}{\bibfnamefont{A.~S.} \bibnamefont{Arnold}},
  \bibinfo{author}{\bibfnamefont{J.~S.} \bibnamefont{Wilson}},
  \bibnamefont{and} \bibinfo{author}{\bibfnamefont{M.~G.}
  \bibnamefont{Boshier}}, \bibinfo{journal}{{Rev.\ Sci.\ Instrum.}}
  \textbf{\bibinfo{volume}{69}}, \bibinfo{pages}{1236} (\bibinfo{year}{1998}).

\bibitem[{\citenamefont{Ricci et~al.}(1995)\citenamefont{Ricci,
  Weidem{\"u}ller, Esslinger, Hemmerich, Zimmermann, Vuletic, K{\"o}nig, and
  H{\"a}nsch}}]{Ricci95}
\bibinfo{author}{\bibfnamefont{L.}~\bibnamefont{Ricci}},
  \bibinfo{author}{\bibfnamefont{M.}~\bibnamefont{Weidem{\"u}ller}},
  \bibinfo{author}{\bibfnamefont{T.}~\bibnamefont{Esslinger}},
  \bibinfo{author}{\bibfnamefont{A.}~\bibnamefont{Hemmerich}},
  \bibinfo{author}{\bibfnamefont{C.}~\bibnamefont{Zimmermann}},
  \bibinfo{author}{\bibfnamefont{V.}~\bibnamefont{Vuletic}},
  \bibinfo{author}{\bibfnamefont{W.}~\bibnamefont{K{\"o}nig}},
  \bibnamefont{and} \bibinfo{author}{\bibfnamefont{T.~W.}
  \bibnamefont{H{\"a}nsch}}, \bibinfo{journal}{{Opt.\ Comm.}}
  \textbf{\bibinfo{volume}{117}}, \bibinfo{pages}{541} (\bibinfo{year}{1995}).

\bibitem[{\citenamefont{Vassiliev et~al.}(2006)\citenamefont{Vassiliev, Zibrov,
  and Velichansky}}]{Vassiliev06}
\bibinfo{author}{\bibfnamefont{V.~V.} \bibnamefont{Vassiliev}},
  \bibinfo{author}{\bibfnamefont{S.~A.} \bibnamefont{Zibrov}},
  \bibnamefont{and} \bibinfo{author}{\bibfnamefont{V.~L.}
  \bibnamefont{Velichansky}}, \bibinfo{journal}{{Rev.\ Sci.\ Instrum.}}
  \textbf{\bibinfo{volume}{77}}, \bibinfo{pages}{013102}
  (\bibinfo{year}{2006}).

\bibitem[{\citenamefont{Zorabedian and Trutn{a, Jr.}}(1988)}]{Zorabedian88}
\bibinfo{author}{\bibfnamefont{P.}~\bibnamefont{Zorabedian}} \bibnamefont{and}
  \bibinfo{author}{\bibfnamefont{W.~R.} \bibnamefont{Trutn{a, Jr.}}},
  \bibinfo{journal}{{Opt.\ Lett.}} \textbf{\bibinfo{volume}{13}},
  \bibinfo{pages}{826} (\bibinfo{year}{1988}).

\bibitem[{\citenamefont{Baillard et~al.}(2006)\citenamefont{Baillard, Gauguet,
  Bize, Lemonde, Laurent, Clairon, and Rosenbusch}}]{Baillard06}
\bibinfo{author}{\bibfnamefont{X.}~\bibnamefont{Baillard}},
  \bibinfo{author}{\bibfnamefont{A.}~\bibnamefont{Gauguet}},
  \bibinfo{author}{\bibfnamefont{S.}~\bibnamefont{Bize}},
  \bibinfo{author}{\bibfnamefont{P.}~\bibnamefont{Lemonde}},
  \bibinfo{author}{\bibfnamefont{P.}~\bibnamefont{Laurent}},
  \bibinfo{author}{\bibfnamefont{A.}~\bibnamefont{Clairon}}, \bibnamefont{and}
  \bibinfo{author}{\bibfnamefont{P.}~\bibnamefont{Rosenbusch}},
  \bibinfo{journal}{{Opt.\ Comm.}} \textbf{\bibinfo{volume}{266}},
  \bibinfo{pages}{609} (\bibinfo{year}{2006}).

\bibitem[{\citenamefont{Papp}(2007)}]{Papp07thesis}
\bibinfo{author}{\bibfnamefont{S.~B.} \bibnamefont{Papp}},
  \bibinfo{type}{P{h.D.} thesis}, \bibinfo{school}{University of Colorado
  Boulder} (\bibinfo{year}{2007}).

\bibitem[{\citenamefont{Meyrath}(2003)}]{Meyrath03notes}
\bibinfo{author}{\bibfnamefont{T.~P.} \bibnamefont{Meyrath}},
  \emph{\bibinfo{title}{An analog current controller design for laser diodes}},
  \bibinfo{howpublished}{available online at
  \url{http://george.ph.utexas.edu/~meyrath/informal}} (\bibinfo{year}{2003}).

\bibitem[{\citenamefont{Steck}(2001)}]{Steck01}
\bibinfo{author}{\bibfnamefont{D.~A.} \bibnamefont{Steck}},
  \bibinfo{type}{P{h.D.} thesis}, \bibinfo{school}{University of Texas Austin}
  (\bibinfo{year}{2001}).

\bibitem[{\citenamefont{McNicholl and Metcalf}(1985)}]{McNicholl85}
\bibinfo{author}{\bibfnamefont{P.}~\bibnamefont{McNicholl}} \bibnamefont{and}
  \bibinfo{author}{\bibfnamefont{J.~H.} \bibnamefont{Metcalf}},
  \bibinfo{journal}{{Appl.\ Opt.}} \textbf{\bibinfo{volume}{24}},
  \bibinfo{pages}{2757} (\bibinfo{year}{1985}).

\bibitem[{\citenamefont{Libbrecht and Hall}(1993)}]{Libbrecht93}
\bibinfo{author}{\bibfnamefont{K.~G.} \bibnamefont{Libbrecht}}
  \bibnamefont{and} \bibinfo{author}{\bibfnamefont{J.~L.} \bibnamefont{Hall}},
  \bibinfo{journal}{{Rev.\ Sci.\ Instrum.}} \textbf{\bibinfo{volume}{64}},
  \bibinfo{pages}{2133} (\bibinfo{year}{1993}).

\bibitem[{\citenamefont{Gallion et~al.}(1982)\citenamefont{Gallion, Mendieta,
  and Leconte}}]{Gallion82}
\bibinfo{author}{\bibfnamefont{P.}~\bibnamefont{Gallion}},
  \bibinfo{author}{\bibfnamefont{F.~J.} \bibnamefont{Mendieta}},
  \bibnamefont{and} \bibinfo{author}{\bibfnamefont{R.}~\bibnamefont{Leconte}},
  \bibinfo{journal}{{J.\ Opt.\ Soc.\ Am.}} \textbf{\bibinfo{volume}{72}},
  \bibinfo{pages}{1167} (\bibinfo{year}{1982}).

\bibitem[{\citenamefont{Gallion and Debarge}(1984)}]{Gallion84}
\bibinfo{author}{\bibfnamefont{P.}~\bibnamefont{Gallion}} \bibnamefont{and}
  \bibinfo{author}{\bibfnamefont{G.}~\bibnamefont{Debarge}},
  \bibinfo{journal}{{IEEE J.\ Quant.\ Elect.}}
  \textbf{\bibinfo{volume}{QE-20}}, \bibinfo{pages}{343}
  (\bibinfo{year}{1984}).

\bibitem[{\citenamefont{Mercer}(1991)}]{Mercer91}
\bibinfo{author}{\bibfnamefont{L.~B.} \bibnamefont{Mercer}},
  \bibinfo{journal}{{J.\ Lightwave Tech.}} \textbf{\bibinfo{volume}{9}},
  \bibinfo{pages}{485} (\bibinfo{year}{1991}).

\bibitem[{\citenamefont{Ludvigsen et~al.}(1998)\citenamefont{Ludvigsen,
  Tossavainen, and Kaivola}}]{Ludvigsen98}
\bibinfo{author}{\bibfnamefont{H.}~\bibnamefont{Ludvigsen}},
  \bibinfo{author}{\bibfnamefont{M.}~\bibnamefont{Tossavainen}},
  \bibnamefont{and} \bibinfo{author}{\bibfnamefont{M.}~\bibnamefont{Kaivola}},
  \bibinfo{journal}{{Opt.\ Comm.}} \textbf{\bibinfo{volume}{155}},
  \bibinfo{pages}{180} (\bibinfo{year}{1998}).

\bibitem[{\citenamefont{Cutler and Searle}(1966)}]{Cutler66}
\bibinfo{author}{\bibfnamefont{L.~S.} \bibnamefont{Cutler}} \bibnamefont{and}
  \bibinfo{author}{\bibfnamefont{C.}~\bibnamefont{Searle}},
  \bibinfo{journal}{{Proc.\ IEEE}} \textbf{\bibinfo{volume}{54}},
  \bibinfo{pages}{136} (\bibinfo{year}{1966}).

\bibitem[{\citenamefont{Steck}(2006)}]{Steck06notes}
\bibinfo{author}{\bibfnamefont{D.~A.} \bibnamefont{Steck}},
  \emph{\bibinfo{title}{Quantum and atom optics}},
  \bibinfo{howpublished}{course notes available online at
  \url{http://steck.us/teaching}} (\bibinfo{year}{2006}).

\bibitem[{\citenamefont{Schawlow and Townes}(1958)}]{Schawlow58}
\bibinfo{author}{\bibfnamefont{A.~L.} \bibnamefont{Schawlow}} \bibnamefont{and}
  \bibinfo{author}{\bibfnamefont{C.~H.} \bibnamefont{Townes}},
  \bibinfo{journal}{{Phys.\ Rev.}} \textbf{\bibinfo{volume}{112}},
  \bibinfo{pages}{1940} (\bibinfo{year}{1958}).

\bibitem[{\citenamefont{Kazarinov and Henry}(1987)}]{Kazarinov87}
\bibinfo{author}{\bibfnamefont{R.~F.} \bibnamefont{Kazarinov}}
  \bibnamefont{and} \bibinfo{author}{\bibfnamefont{C.~H.} \bibnamefont{Henry}},
  \bibinfo{journal}{{IEEE J.\ Quant.\ Elect.}}
  \textbf{\bibinfo{volume}{QE-23}}, \bibinfo{pages}{1401}
  (\bibinfo{year}{1987}).

\bibitem[{\citenamefont{Henry}(1982)}]{Henry82}
\bibinfo{author}{\bibfnamefont{C.~H.} \bibnamefont{Henry}},
  \bibinfo{journal}{{IEEE\ J.\ of\ Quant.\ Elec.}}
  \textbf{\bibinfo{volume}{QE-18}}, \bibinfo{pages}{259}
  (\bibinfo{year}{1982}).

\bibitem[{\citenamefont{Loh et~al.}(2006)\citenamefont{Loh, Lin, Teper, Cetina,
  Simon, and Vuleti\'{c}}}]{Loh06}
\bibinfo{author}{\bibfnamefont{H.}~\bibnamefont{Loh}},
  \bibinfo{author}{\bibfnamefont{Y.}~\bibnamefont{Lin}},
  \bibinfo{author}{\bibfnamefont{I.}~\bibnamefont{Teper}},
  \bibinfo{author}{\bibfnamefont{M.}~\bibnamefont{Cetina}},
  \bibinfo{author}{\bibfnamefont{J.}~\bibnamefont{Simon}}, \bibnamefont{and}
  \bibinfo{author}{\bibfnamefont{V.}~\bibnamefont{Vuleti\'{c}}},
  \bibinfo{journal}{{Appl.\ Opt.}} \textbf{\bibinfo{volume}{45}},
  \bibinfo{pages}{9191} (\bibinfo{year}{2006}).

\bibitem[{\citenamefont{Saliba and Scholten}(2009)}]{Saliba09}
\bibinfo{author}{\bibfnamefont{S.~D.} \bibnamefont{Saliba}} \bibnamefont{and}
  \bibinfo{author}{\bibfnamefont{R.~E.} \bibnamefont{Scholten}},
  \bibinfo{journal}{{Appl.\ Opt.}} \textbf{\bibinfo{volume}{48}},
  \bibinfo{pages}{6961} (\bibinfo{year}{2009}).

\bibitem[{New(Promotional brochure 2009)}]{NewFocus11brochure}
\emph{\bibinfo{title}{New {F}ocus: {TLB-6900} {V}ortex {II} tunable lasers}}
  (\bibinfo{year}{Promotional brochure 2009}).

\bibitem[{Top(Promotional brochure 2011/2012)}]{Toptica11brochure}
\emph{\bibinfo{title}{Toptica {P}hotonics: {T}unable diode lasers}}
  (\bibinfo{year}{Promotional brochure 2011/2012}).

\bibitem[{Top(Catalog 2011/2012)}]{Toptica11cat}
\emph{\bibinfo{title}{Toptica {P}hotonics: {L}asers for scientific challenges}}
  (\bibinfo{year}{Catalog 2011/2012}).

\bibitem[{Ves(Promotional flyer 2011)}]{VescentCECL}
\emph{\bibinfo{title}{Vescent {P}hotonics: {C}hip {E}xternal-{C}avity {L}aser
  {(CECL)}}} (\bibinfo{year}{Promotional flyer 2011}).

\bibitem[{Ves(2012)}]{VescentCECL2}
\bibinfo{howpublished}{\url{http://www.vescent.com/technology/short-cavity-lasers/}}
  (\bibinfo{year}{2012}).

\bibitem[{Sac(Promotional flyer 2011)}]{SacherLynx}
\emph{\bibinfo{title}{Sacher {L}asertechnik {G}roup: Lynx}}
  (\bibinfo{year}{Promotional flyer 2011}).

\bibitem[{Uni(2011)}]{Unilaser_web}
\bibinfo{howpublished}{\url{http://atomoptics.uoregon.edu/unilaser}}
  (\bibinfo{year}{2011}).

\end{thebibliography}
\end{document}